\documentclass[aps,prd,twocolumn,showpacs,eqsecnum,A4,amsmath,amssymb,nofootinbib]{revtex4-1}

\usepackage{amsfonts}
\usepackage{amsmath}
\usepackage{amssymb}
\usepackage{graphicx}
\usepackage{xcolor}
\usepackage{appendix}

\newcommand{\dd}{{\rm{d}}} 

\newcommand{\im} {\mathrm{i}}

\newcommand{\con}{k}     
\newcommand{\M}{\mu}     

\begin{document}

\title{Kerr-Newman and genuine Kerr black holes in the Bertotti-Robinson magnetic field}

\author{Hryhorii Ovcharenko}
\email{hryhorii.ovcharenko@matfyz.cuni.cz}
\affiliation{Charles University, Faculty of Mathematics and Physics,
Institute of Theoretical Physics,
V~Hole\v{s}ovi\v{c}k\'ach 2, 18000 Prague 8, Czechia}

\author{Ji\v{r}\'i Podolsk\'{y}}
\email{jiri.podolsky@matfyz.cuni.cz}
\affiliation{Charles University, Faculty of Mathematics and Physics,
Institute of Theoretical Physics,
V~Hole\v{s}ovi\v{c}k\'ach 2, 18000 Prague 8, Czechia}

\begin{abstract}
We present a spacetime that describes a rotating and charged black hole immersed in an external magnetic (or electric) field which is oriented along the rotational axis. This solution is characterized by four physical parameters, namely the black hole mass $m$, its charge $e$, Kerr-like rotation $a$, and the value of the external field $B$. Such exact solution to the Einstein-Maxwell equations has a surprisingly simple form. Moreover, for $B=0$ it reduces to standard Kerr-Newman metric, while for $m=0$ and $e=0$ we recover the Bertotti-Robinson universe with a uniform (electro-)magnetic field. We also show that the Kerr-BR solution presented in [Phys. Rev. Lett. {\bf 135}, 181401 (2025)] has a special charge, and we identify a genuine uncharged Kerr black hole immersed in the magnetic field. In the weak-field limit, it gives the Wald solution. We analyze the main physical properties of the new family, namely singularities, horizons, regularity of axes, charges, entropy, and temperature.
\end{abstract}

\date{\today}



\maketitle


\section{Introduction}

Soon after Einstein formulated general relativity, an idea of gravitationally collapsed singular ``objects'' emerged. But it took many decades to prove the existence of black holes observationally, notably by studying specific astrophysical X-ray sources, measuring high velocities of S-stars surrounding Sgr~A* at the center of Galaxy, detecting gravitational waves from (hundreds of) binary black hole coalescences, and even taking the first images of a shadow of black holes in M87* and Sgr~A*.

Exact models of black holes are also known for a century. The most important are the Schwarzschild (1916), Reissner-Nordstr\"{o}m (1916), Kerr (1963), and Kerr-Newman solutions (1965). They are the unique spherical and axisymmetric stationary electro-vacuum spacetimes that are asymptotically flat, and they describe rotating and charged black holes \cite{Stephani:2003tm, GriffithsPodolsky:2009}. Actually, they belong to a larger class of spacetimes of algebraic type~D and double-aligned electromagnetic field,\footnote{By the (non-)alignment we mean here the (non-)alignment of the eigen-directions of the Faraday tensor and the principal null directions of the Weyl tensor. It does not refer to the geometric relation of the external field to the axis of rotation. These coincide for the whole Kerr-Newman-Bertotti-Robinson family.} with additional physical parameters representing a cosmological constant, acceleration, and NUT charge \cite{Debever1971,Plebanski1976,Debever1984,Plebanski1976,GRIFFITHS2006,Griffiths2005,PodolskyGriffiths:2006,PodolskyVratny:2021,PodolskyVratny:2023,Astorino:2024b, OvcharenkoPodolskyAstorino:2025a,OvcharenkoPodolskyAstorino:2025b}.

Recently, we found a novel family of black holes that are also exact type D solutions to the Einstein-Maxwell equations, but their electromagnetic field is \emph{not} aligned \cite{OvcharenkoPodolsky:2025}. Among its subclasses we identified black holes which are not asymptotically flat, but are immersed in the  electromagnetic field of the uniform Bertotti-Robinson (BR) universe. Specific interaction between the charged and rotating (possibly accelerating) black hole and the external BR field causes various effects, changing the position and shape of the horizons, modifying  geodesic trajectories, exhibiting the Meissner effect, etc. We presented such Kerr-BR black holes in \cite{PodolskyOvcharenko:2025}, and its non-rotating  Reissner-Nordstr\"om-BR and Schwarzschild-BR subcases in \cite{OvcharenkoPodolsky:2026a}. The works attracted a considerable attention because these exact spacetimes do not have some unpleasant properties of previously known and thoroughly investigated class of Kerr-Newman-Melvin black holes of algebraic type~I found by Ernst and Wild \cite{Ernst1976_2, Ernst1976_3} with the Bonnor-Melvin \cite{Bonor:1954, Melvin:1964} magnetic field, see e.g. the reviews \cite{GibbonsMujtabaPope:2013, GibbonsPangPope:2014} and recent works \cite{DiPintoKlemmVigano:2025,Astorino:2025,TaylorRitz:2025a,TaylorRitz:2025b}.

However, in the studies of our new family of solutions many questions have remained open. One of them was the proper identification of the physical charge of the black holes, which is not a straightforward task due to the non-linear interaction between the external BR field and the mass, charge, rotation and acceleration of the black hole. The purpose of this paper is to solve this problem. Indeed, we identify the Kerr-Newman-Bertotti-Robinson (KN-BR) spacetime, and the genuine Kerr-BR$_0$ black hole \emph{without charge}, distinct from the Kerr-BR solution.

This new class of KN-BR black holes is presented in Sec.~\ref{Sec.KNBR}, in Sec.~\ref{Sec.subcases} we describe all its subcases, and in Sec.~\ref{Sec.physics} we investigate the key physical properties.

\section{Kerr-Newman-Bertotti-Robinson spacetime}
\label{Sec.KNBR}

Metric of the Kerr-Newman black hole immersed in the Bertotti-Robinson universe filled with a uniform electromagnetic field (our new exact KN-BR spacetime)~is
\begin{align}
\dd s^2&= \dfrac{1}{\Omega^2}\Big[
    - \dfrac{Q}{\rho^2}\big(\dd t-a \sin^2\theta\, \dd \varphi\big)^2
    + \dfrac{\rho^2}{Q}\,\dd r^2
    + \dfrac{\rho^2}{P}\,\dd\theta^2 \nonumber\\
    &\hspace{12mm}
    +\dfrac{P}{\rho^2} \sin^2\theta \big(a\dd t-(r^2+a^2)\dd \varphi\big)^2\Big].\label{metr}
\end{align}
The metric functions are the following simple (combinations of) quadratic expressions in $r$ and $\cos\theta$,
\begin{align}
    \rho^2 &= r^2+a^2\cos^2\theta \,, \nonumber\\
    P &= 1 + B^2 \M^2 \cos^2\theta \,, \nonumber\\
    Q &= I\,\Delta\,, \nonumber\\
    \Omega^2 &= I-B^2 \Delta \cos^2\theta\,, \nonumber\\
    I &= (1 + \con\,B\,r)^2+B^2 r^2\,, \nonumber\\
    \Delta   &= (1+\con^2)\,a^2 -2m\,r + (1 + \con^2 - e^2/a^2)\, r^2\,, \label{Delta_eq}
\end{align}
with the constants
\begin{align}
    \con   &= \dfrac{e\, s - a\, m\, B}{a\,(1+e^2 B^2)}\,,\label{c_eq}\\
    s^2 &= 1+(e^2-a^2)B^2-(m^2+e^2)\,a^2 B^4\,,\label{s_eq}\\
    \M^2 &= m^2-(1+\con^2)^2 a^2 +(1+\con^2)\,e^2\,. \label{M_eq}
\end{align}
We assume ${\M\ge0}$, and consider only the square root ${s\ge0}$ of \eqref{s_eq} (the case ${s<0}$ is included in \eqref{c_eq} if we admit both signs of the charge, ${e>0}$ and ${e<0}$).

The range of the angular coordinates is ${\theta\in[0,\pi]}$ and ${\varphi\in [0,2\pi C)}$, where the constant $C$ is given by Eq.~\eqref{conicity}. By redefining $\varphi$ to ${\phi = (1+B^2\M^2)\,\varphi}$, the new coordinate $\phi$ has the usual range ${\phi\in [0,2\pi)}$.

The Maxwell field is determined by the potential
\begin{align}
\mathbf{A}&=\  \dfrac{1}{\Omega}\Big[
    -(1+\con^2)B+(mB\cos^2\theta-\con)\,\frac{r}{\rho^2}\Big]\,a\,\dd t \nonumber\\
    &\ \, +\dfrac{1}{\Omega}\Big[ (1+\con^2)\,a^2B + \big( -mB (r^2+a^2)\cos^2\theta \nonumber\\
   &\qquad\ +\con\,a^2\sin^2\theta \big)\frac{r}{\rho^2} -\con\,r -\frac{1}{B}\,\Big]\dd\varphi
   + \mathbf{A}_0\,,\label{A_vec-explicitly}
\end{align}
where $\mathbf{A}_0$ is a closed 1-form representing the gauge freedom. We obtained this class of solutions to the Einstein-Maxwell equations by performing a specific limit of the general type D spacetimes with non-aligned electromagnetic field that we found in \cite{OvcharenkoPodolsky:2025}.

This class of black holes depends on 4 physical parameters, namely the

- \emph{mass} parameter $m$,

- \emph{rotation} parameter $a$,

- \emph{external magnetic field} parameter $B$,

- \emph{electric charge of the black hole} given by $e$.

We will prove below that the parameter $e$ is the \emph{proper charge of the black hole itself}, namely that ${ q_e =e\, C}$, where $C$ is the conicity parameter \eqref{conicity} that makes \emph{both
axes ${\theta=0}$ and ${\theta=\pi}$ regular}. Let us emphasize that these black hole spacetimes do \emph{not} contain any cosmic string (deficit angle).  This can be  immediately seen from the metric function~$P$ given by \eqref{Delta_eq}. It contains only the \emph{quadratic} term $\cos^2\theta$, so that both the axes can be regularized simultaneously.

\newpage

\section{Main subcases}
\label{Sec.subcases}

\begin{figure}[h!]
\vspace{0mm}
    \centering
 \includegraphics[width=1\linewidth, trim={1mm 0 1.5mm 0},clip]{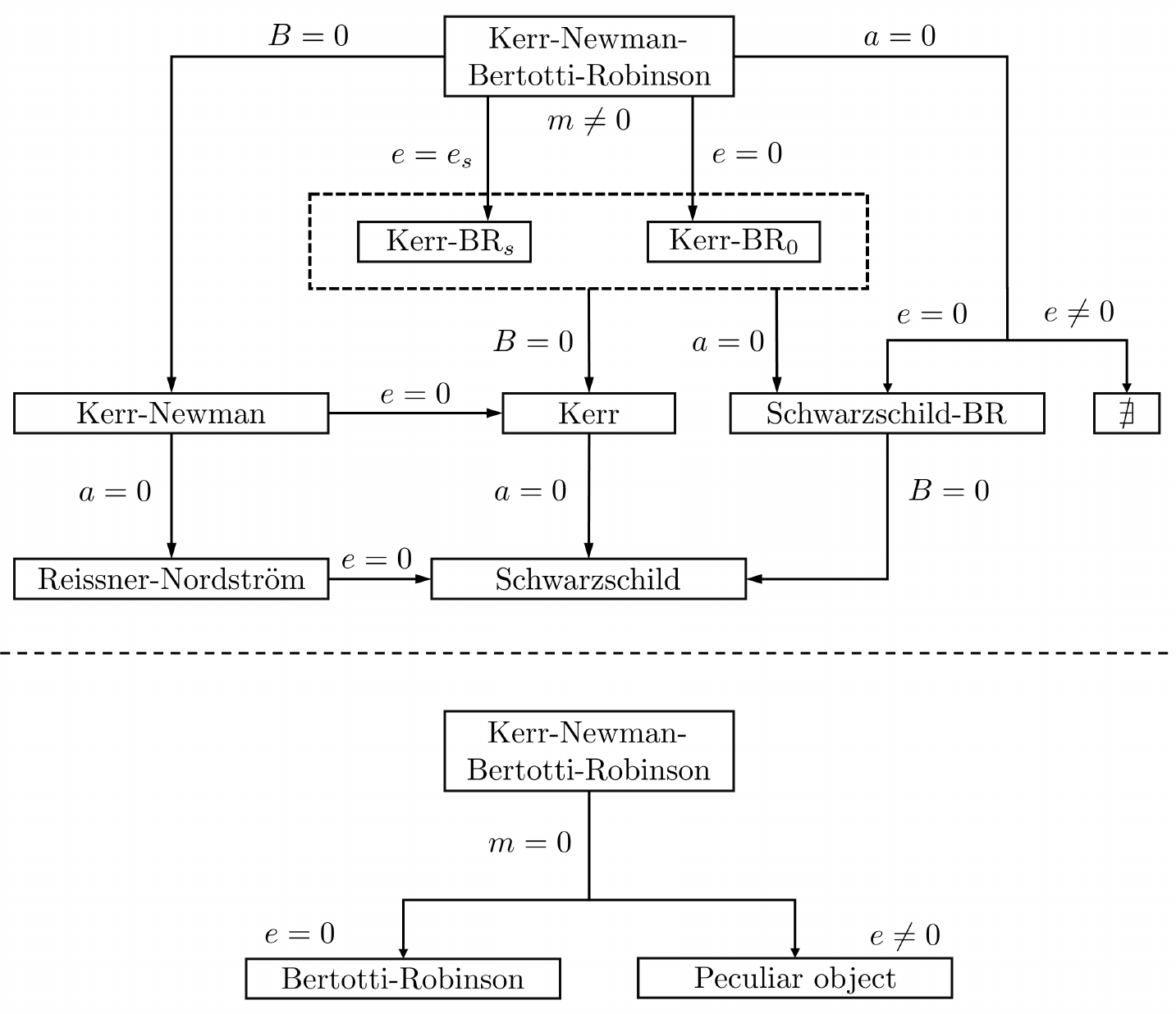}
    \caption{A scheme describing various subcases of the new class of Kerr-Newman-Bertotti-Robinson spacetimes. The special value of the electric charge is ${e_s=m a B/\sqrt{1-a^2B^2}}$. The upper part describes black holes with ${m\ne0}$, while the lower part corresponds to ${m=0}$.}
    \label{scheme}
\end{figure}

Let us now discuss various possible subcases of the whole family \eqref{metr}--\eqref{A_vec-explicitly}. All these important metrics, obtained simply by setting the corresponding physical parameter to zero, are summarized in the scheme in Fig.~\ref{scheme}.

\subsection{Kerr-Newman: No magnetic field (${B=0}$)}

The coefficients \eqref{s_eq}, \eqref{c_eq} simplify to
${s=1}$, ${\con=e/a}$, so that ${I=1}$ and the  metric functions \eqref{Delta_eq} become
\begin{align}\label{KN}
    P=1\,,\qquad
    Q=r^2-2m r+a^2+e^2\,,\qquad
    \Omega=1\,.
\end{align}
The metric \eqref{metr} is thus the \emph{standard form of the Kerr-Newman spacetime} in spheroidal (Boyer-Lindquist) coordinates, that represents rotating charged black hole.

The potential \eqref{A_vec-explicitly} of the Maxwell field reduces to
\begin{align}\label{A-Kerr-Newman}
    \mathbf{A}=-\dfrac{e \,r}{\rho^2}(\dd t-a \sin^2\theta\, \dd \varphi)
       -\dfrac{1}{B}\,\dd \varphi+\mathbf{A}_0\,,
\end{align}
where we used
${\displaystyle\lim_{B\to 0}\frac{1}{\Omega\,B} = \lim_{B\to 0}\frac{1}{B} - \con\,r}$. Choosing
\begin{align}\label{gauge A0}
\mathbf{A}_0 = \dfrac{1}{B}\,\dd \varphi\,,
\end{align}
the 4-potential \eqref{A-Kerr-Newman} takes the usual form for the Kerr-Newman spacetime \cite{Stephani:2003tm,GriffithsPodolsky:2009}.

\subsection{Kerr-BR$_s$ spacetime with charge (${e=e_s\ne0}$)}

Let us investigate the special case that gives the Kerr-BR spacetime found in \cite{PodolskyOvcharenko:2025}. One would intuitively expect that this corresponds to the simplest choice of the charge ${e=0}$, but \emph{it is not so}. By calculating the electric flux through the horizon for this spacetime (see Sec.~\ref{charges}), one finds that it is non-zero. Actually, the Kerr-BR$_s$ spacetime of \cite{PodolskyOvcharenko:2025} is obtained from the general Kerr-Newman-BR family \eqref{metr} by choosing a very \emph{special} value of~$e$, namely
\begin{align}
    e_s = m\,\dfrac{a  B}{\sqrt{1-a^2B^2}}\,. \label{especial}
\end{align}
For such a \emph{unique non-zero} value of the charge $e$, the key constant $\con$ given by \eqref{c_eq} is zero,
\begin{align}
\con=0\,,
\end{align}
because ${s=\sqrt{1-a^2B^2}}$. It leads to a significant simplification of the metric functions,
\begin{align}
    P&=  1+B^2\Big(\dfrac{m^2}{1-a^2 B^2}-a^2\Big)\cos^2\theta\,, \nonumber\\
    Q&=  (1+B^2r^2)\,\Delta\,,\nonumber\\
    \Delta&=  a^2-2m\, r+\Big(1-\dfrac{m^2B^2}{1-a^2 B^2}\Big)r^2\,,\nonumber\\
    \Omega^2&= (1+B^2r^2)-B^2\Delta \cos^2\theta\,. \label{Omega-Kerr-BR}
\end{align}
By a simple  re-definition of the mass parameter~$m$ to
\begin{align}
    \tilde{m} \equiv m\,\dfrac{I_1}{I_2}\,,\quad
    I_1=1-\frac{1}{2}a^2 B^2\,,\quad
    I_2=1-a^2 B^2\,,\label{mp}
\end{align}
the metric functions $P$ and $\Delta$ become
\begin{align}
    P&=1+B^2\Big(\tilde{m}^2\,\dfrac{I_2}{I_1^2}-a^2\Big)\cos^2\theta\,,\nonumber\\
    \Delta&=a^2-2\tilde{m}\,\dfrac{I_2}{I_1} \,r
        +\Big(1-B^2\tilde{m}^2\,\dfrac{I_2}{I_1^2}\Big)\,r^2\,.
\end{align}
That is exactly the Kerr-BR spacetime in the form presented in \cite{PodolskyOvcharenko:2025}, which we \emph{now rename} to Kerr-BR$_s$.

Notice that for ${B=0}$ or ${a=0}$, we get ${\tilde{m}=m}$. In such cases ${e_s=0}$, which gives black holes \emph{without any charge}, i.e. that of Kerr or Schwarzschild-BR, respectively.  Moreover, for ${aB}$ small we obtain ${\tilde{m} \approx m \, (1  + \tfrac{1}{2}\,a^2B^2)}$, i.e. the mass parameters $m$ and $\tilde{m}$ agree up to \emph{linear} terms in $a$ and $B$.

\subsection{Genuine Kerr-BR$_0$ spacetime: No charge (${e=0}$)}

In the case of \emph{black holes with no electric charge}, ${e=0}$, the constants \eqref{c_eq}, \eqref{s_eq} reduce to
\begin{align}
    \con = - m B\,,\qquad
    s^2  = 1-a^2 B^2(1-m^2 B^2)\,.
\end{align}
Then the metric functions become
\begin{align}
    P&= 1+B^2 \big[\,m^2-a^2(1+m^2B^2)^2\big]\cos^2\theta\,, \nonumber\\
    Q&= I\,\Delta\,, \nonumber\\
    \Omega^2&= I-B^2\Delta \cos^2\theta\,,\nonumber\\
    I&= (1-m B^2 r)^2+B^2r^2\,,\nonumber\\
    \Delta&= (r^2+a^2)(1+m^2 B^2)-2 m\,r\,. \label{Delta-Kerr-BR0}
\end{align}
The Maxwell-field potential \eqref{A_vec-explicitly} with  \eqref{gauge A0} becomes
\begin{align}
\mathbf{A}=& -  \dfrac{B}{\Omega}\,\Big[\,
    (1+m^2 B^2) - m\,r\,\frac{1+\cos^2\theta}{r^2+a^2\cos^2\theta}\,\Big]\,a\,\dd t \nonumber\\
    & +\dfrac{B}{\Omega}\,\Big[ a^2 (1+m^2 B^2)
      - m\,r\,\frac{(a^2-r^2) \sin^2\theta}{r^2+a^2\cos^2\theta}\,\Big]\dd\varphi \nonumber\\
    & + \frac{\Omega-1}{\Omega\,B}\,\dd\varphi\,.
   \label{A_vec-explicitly-Kerr-BR0}
\end{align}
The last term vanishes in the limit ${B \to 0}$ because ${(\Omega-1)\propto B^2}$ for small $B$, and ${\Omega \to 1}$. The potential $\mathbf{A}$ thus vanishes for ${B=0}$ because there is no charge of the black hole, and also there is no external Bertotti-Robinson magnetic field.

Let us emphasize again that this spacetime \emph{is different from the Kerr-BR$_s$ spacetime} \eqref{Omega-Kerr-BR} presented in \cite{PodolskyOvcharenko:2025}. To avoid a confusion, we denote the spacetime \eqref{Delta-Kerr-BR0} as the \emph{Kerr-BR$_0$ spacetime}, in which the index~$0$ indicates that the physical electric charge of such a spacetime is zero, ${e=0}$. The Kerr-BR$_s$ spacetime has ${e\ne0}$ given by \eqref{especial}, albeit negligible for small ${aB}$. Thus, the  Kerr-BR$_0$ spacetime is the \emph{genuine} rotating Kerr black hole immersed in the external Bertotti-Robinson magnetic field.

\subsection{Schwarzschild-BR spacetime: No charge and no rotation (${e=0=a}$)}\label{sec_Schw_BR_subcase}

Interestingly, in the ${a\to 0}$ limit \emph{both} Kerr-BR$_s$ and Kerr-BR$_0$ spacetimes give the Schwarzschild-BR non-rotating black hole because the electric charge of the Kerr-BR$_s$ spacetime is ${e_s \approx maB \to 0}$. In such a limit both the rotation and the charge of the black hole vanish, and from  \eqref{metr}--\eqref{s_eq} we obtain
\begin{align}
\dd s^2&= \dfrac{1}{\Omega^2}\Big[
    -\dfrac{Q}{r^2}\,\dd t^2
    +\dfrac{r^2}{Q}\,\dd r^2 + r^2\,\Big(\dfrac{\dd\theta^2}{P} + P\sin^2\theta \,\dd \varphi^2\Big)\Big],\label{metr-Schw-BR}
\end{align}
with the metric functions
\begin{align}
    P&= 1+m^2 B^2 \cos^2\theta\,,\qquad
    \Omega^2= I-B^2\Delta \cos^2\theta\,,\nonumber\\
    Q&= I\,\Delta\,,\qquad
    I = (1-m B^2 r)^2+B^2r^2\,,\nonumber\\
    \Delta &= \big[(1+m^2B^2)\,r - 2m\,\big]\,r\,.
\end{align}
The potential of the external magnetic field becomes
\begin{align}
\mathbf{A} = m B\,\dfrac{r \sin^2\theta}{\Omega}\,\dd\varphi
   + \dfrac{\Omega-1}{\Omega\,B}\,\dd\varphi
   \,,\label{A_vec-explicitly-Schw-BR0}
\end{align}
where the first term vanishes if ${mB=0}$, while the second term vanishes in the ${B \to 0}$ limit. Performing the transformation ${r=r'/(1+m B^2\, r')}$,
the metric functions in \eqref{metr-Schw-BR} become
${\Omega'^2= I'-B^2\Delta' \cos^2\theta}$,
${Q'= I'\,\Delta'}$,
${I'= 1+B^2r'^2}$,
${\Delta'= r'\,\big[r'(1-B^2m^2)-2m\big]}$. This is the original form of the \emph{Schwarzschild-BR spacetime}, as it was introduced in Eq.~(18) of~\cite{PodolskyOvcharenko:2025}.

\subsection{No rotation (${a=0}$): Not possible for ${e\ne0}$}

The limit ${a \to 0}$ with ${e \ne 0}$ leads to diverging metric functions. This agrees with the results of \cite{OvcharenkoPodolsky:2026a}, where it was shown that \emph{there are no static electrically charged black holes of type D in the external Bertotti-Robinson magnetic field}. (For ${e=0}$ or ${B=0}$ in the limit ${a \to 0}$ we get the Schwarzschild-BR or Reissner-Nordstr\"om spacetimes, respectively. The Schwarzschild-BR black hole is also obtained for ${a \to 0}$ with ${\gamma\equiv e/a}$ fixed.)

\subsection{No mass and no charge: Bertotti--Robinson background (${m=0=e}$)}\label{sec_m_q_0}

When there is no black hole (the charge and mass are both zero) then
\begin{align}
    \con&=0\,,\qquad  s^2=1-a^2 B^2\,,\qquad  I=1+B^2r^2\,,\nonumber\\
    \Delta&=r^2+a^2\,,\qquad \Omega^2=P+B^2r^2\sin^2\theta\,, \\
    P&=1-B^2a^2\cos^2\theta\,,\qquad   Q =(1+B^2r^2)(r^2+a^2)\,. \nonumber
\end{align}
These are the metric functions that we derived in Section II.B in~\cite{PodolskyOvcharenko:2025}. Therein, it was shown that by performing the transformation $r,\theta,t,\varphi \mapsto R,\Theta,\tau,\phi$ given by
\begin{align}
    \dfrac{R^2}{E^2}&=P\,\dfrac{1+B^2 r^2}{\Omega^2}-1\,,\qquad
    E \sin\Theta=\dfrac{\sqrt{r^2+a^2}}{A\,\Omega}\sin\theta\,,\nonumber\\
    \tau&=A^{-1}\,t\,,\qquad
    \phi=A^2\varphi+B^2 a \,t\,,\label{BR_transform}
\end{align}
where $A=\sqrt{1-a^2B^2}$ and $E=(AB)^{-1}$, the metric (\ref{metr}) takes the usual form of the \emph{Bertotti--Robinson spacetime}
\begin{align}
    \dd s^2=&-(1+R^2/E^2)\,\dd \tau^2+(1+R^2/E^2)\,\dd R^2 \nonumber\\
      &+E^2(\dd\Theta^2+\sin^2\Theta\,\dd \phi^2)\,.
\end{align}
The potential \eqref{A_vec-explicitly-Schw-BR0} for ${m=0}$ reduces to
\begin{align}
\mathbf{A}&= \dfrac{\Omega-1}{\Omega\,B}\,\dd\varphi
   \,,\label{A_vec-explicitly-BR}
\end{align}
where ${\Omega^2=1+B^2r^2\sin^2\theta}$. This fully agrees with the Bertotti-Robinson magnetic field given by Eq.~(8) in \cite{PodolskyOvcharenko:2025} for ${\gamma=0}$ and ${a=0=m}$.

These observations allow us to interpret the complete spacetime, given by the metric \eqref{metr}--\eqref{A_vec-explicitly}, as the Kerr-Newman black hole immersed in the external magnetic field of the Bertotti-Robinson type.

\subsection{No mass, but with charge and rotation:  peculiar object (${m=0}$)}

In this exceptional case, we get
\begin{align}
    \con^2   &= \dfrac{e^2}{a^2}\dfrac{1-a^2B^2}{1+e^2 B^2}\,,\label{c_eq-m=0}
\end{align}
which puts an upper limit on the magnetic field ${aB\le1}$. It reduces to ${k=e/a}$ in the Kerr-Newman case ${B=0}$. This can be understood as an  ``overcharged'' and/or ``overrotating'' object ``without mass'' in an external magnetic field. It is a specific type~D generalization of the massless (${m=0}$) Reissner-Nordstr\"om and massless Kerr-Newman naked singularity. Lacking the horizons, it is not a black hole, unless we consider a special situation discussed in Sec.~\ref{horizons} which admits horizons~\eqref{horiz-m=0case}.

\subsection{Test field (${B \to 0}$, ${e=0}$):  Wald solution }

Finally, we show that our new exact solution involves the solution found in 1974 by Wald \cite{Wald:1974}. In this seminal work, Wald presented a \emph{test} uniform electromagnetic field in the Kerr background, without considering its backreaction on the geometry. The Faraday tensor is given by Eq.~(3.1) in \cite{Wald:1974} as ${\mathbf{F}_{\rm Wald}=\dd\mathbf{A}_{\rm Wald}=\tfrac{1}{2} B\big(\dd \psi+2 a \,\dd \eta\big)}$,
where ${\psi=\psi_{\mu}\,\dd x^\mu}$ and ${\eta=\eta_{\mu}\,\dd x^\mu}$ are duals to the Killing vectors ${\psi^{\mu}=\partial_t}$ and ${\eta^{\mu}=\partial_{\varphi}}$, respectively. The Wald 4-potential for such a test field can thus be expressed as
$\mathbf{A}_{\rm Wald}=\tfrac{1}{2} B \big[
  \big(  g_{\varphi t}+2 a \,g_{t t}\big)\dd t
  + \big(  g_{\varphi \varphi}+2 a \,g_{t \varphi}\big)\dd \varphi \big]$.
For the background metric \eqref{metr} of the Kerr black hole  (${B=0=e}$), which has the metric coefficients \eqref{KN}, we obtain
\begin{align}
    &\mathbf{A}_{\rm Wald}=  \dfrac{B}{\rho^2}\, a m \,r\, (1+\cos^2\theta)\,\dd t
    +\dfrac{B}{2\rho^2}\big[\,r^4+a^4 \cos^2\theta  \nonumber\\
    &\quad -a^2r\,(2m-r)(1+\cos^2\theta)\big]\sin^2\theta\,\dd \varphi -aB\,\dd t.
\end{align}
Notice that the last term is just a gauge. Now, let us take the complete potential \eqref{A_vec-explicitly-Kerr-BR0} of our genuine Kerr-BR$_0$ spacetime having no electric charge (${e=0}$), and perform its \emph{linearization} for small values of $B$, yielding
\begin{align}
    \mathbf{A}_{\rm linearized}=  \mathbf{A}_{\rm Wald} - \tfrac{1}{2}B a^2\dd \varphi\,.
\end{align}
Up to the gauge term they are the same. The Wald solution is thus a weak-field limit of our exact spacetime.

\section{Physical discussion of the Kerr-Newman-BR spacetime}
\label{Sec.physics}

Now we will study main physical aspects of the general family of Kerr-Newman-Bertotti-Robinson black holes.

\subsection{Curvature singularities}\label{sec_sing}

In the canonical null complex tetrad
\begin{align}
    \mathbf{k}&=\dfrac{\Omega}{\rho\sqrt{2Q}}\Big[(r^2+a^2)\,\partial_{t}
         +a \,\partial_{\varphi} + Q\,\partial_r\,\Big],\nonumber\\
    \mathbf{l}&=\dfrac{\Omega}{\rho\sqrt{2Q}}\Big[(r^2+a^2)\,\partial_{t}
         +a \,\partial_{\varphi} - Q\,\partial_r\,\Big],\\
   \mathbf{m}&=\dfrac{\Omega}{\rho\sqrt{2P}}\Big[ a \sin\theta\,\partial_{t}
         + \dfrac{1}{\sin\theta}\,\partial_{\varphi} + \im \,P\,\partial_{\theta}\,\Big],\nonumber
\end{align}
the only component of the Weyl tensor is
\begin{align}\label{Psi2}
    \Psi_2&=\dfrac{1}{12}\dfrac{\Omega^2}{\rho^2}\,(r+\im\,a\, x)^3
     \nonumber\\
      & \quad \times\Big[\Big(\frac{Q}{(r+\im\,a\, x)^3}\Big)_{,rr}+\Big(\frac{(1-x^2)P}{(r+\im\,a\, x)^3}\Big)_{,xx}\,
    \Big],
\end{align}
in which ${x=\cos\theta}$, so that ${\partial_\theta=-\sin\theta\,\partial_x}$. This clearly exhibits that the whole family is of algebraic type~D.

Moreover, we see that there is a curvature singularity at ${\rho=0}$. In view of \eqref{Delta_eq} this requires both ${r=0}$ and ${\theta=\pi/2}$ (unless ${a=0}$, in which case the divergence is at ${r=0}$ for any~$\theta$). These rotating charged black holes in the magnetic field thus contain a \emph{ring-like curvature singularity}, analogous to the Kerr black hole one.

From \eqref{Psi2} we also conclude that the spacetimes become conformally flat at ${\Omega=0}$.  The specific global structure is investigated in \cite{OP-prepar}. Let us also remark that for ${B=0}$ the general form \eqref{Psi2} directly reduces to
\begin{align}\label{Psi2-B=0}
    \Psi_2 = \dfrac{1}{(r+\im\,a\, x)^3}
    \Big[-m + \frac{e^2}{r-\im\,a\, x}\, \Big],
\end{align}
which is the usual expression for Kerr-Newman black hole. For ${a=0}$ it simplifies to
${ \Psi_2 = -m/r^{3} + e^2/r^{4}}$, valid for the Reissner-Nordstr\"om black hole \cite{GriffithsPodolsky:2009}. These are asymptotically flat as ${r \to \infty}$.


\subsection{Horizons}
\label{horizons}

Horizons of the spacetime given by the metric \eqref{metr} are null hypersurfaces  ${r=r_h}$ such that ${Q(r_h)=0}$. Crossing this hypersurface, the  ``radial'' coordinate $r$ changes its character from spatial (for ${Q>0}$) to timelike (for ${Q<0}$), and simultaneously the first term in the metric indicating the timelike coordinate becomes spatial. For ${a=0}$ the Killing vector $\partial_t$ becomes null, so that $r_h$ localizes the Killing horizon. Because for the Kerr-Newman-BR spacetime the metric function $Q$ given by \eqref{Delta_eq} is factorized to ${Q=I\,\Delta}$, and ${I>0}$ everywhere, it is easy to determine all the horizons by simply solving the quadratic equation ${\Delta(r_h)=0}$, which in general has two roots
\begin{align}\label{general-horizons}
    r_{\pm} =\dfrac{m\pm \M} {1+\con^2-e^2/a^2}\,.
\end{align}

\emph{Extremal} horizon located at $r_{\rm e}=m/(1+\con^2-e^2/a^2)\equiv a^2(1+k^2)/m$ occurs if the physical parameters $m, a, e, B$ satisfy the condition ${\M=0}$, which using \eqref{M_eq} is
\begin{align}\label{extremality}
    m^2 = a^2 (1+\con^2)^2 - e^2 (1+\con^2)\,,
\end{align}
where $k$ is given by \eqref{c_eq}. Generally, it is not easy to solve this equation because it involves $s$, given by the square root of non-trivial combinations of the physical parameters \eqref{s_eq}. We thus only analyze the special cases.

\begin{itemize}

\item In the ${B=0}$ subcase (no external magnetic field) there is ${k=e/a}$, and we immediately obtain ${r_{\pm}=m\pm \sqrt{m^2-a^2-e^2}}$, as expected for the Kerr-Newman family of black holes \cite{Stephani:2003tm, GriffithsPodolsky:2009}.

\item In the case ${e=e_s}$, we obtain the expressions for the horizons positions in the Kerr-BR$_s$ charged black hole, investigated in detail in \cite{PodolskyOvcharenko:2025}.
        Notice that in such a case ${k=0}$, and the extremality condition \eqref{extremality} simplifies to ${m^2 = a^2 - e_s^2}$, i.e., ${e_s=a^2B}$.

\item In the case ${e=0}$ of \emph{genuine uncharged} Kerr-BR$_0$ spacetime, the key constant is ${\con = - m B}$, so that the two horizons are located at
\begin{align}
    r_{\pm}=\dfrac{m\pm \sqrt{m^2-a^2(1+m^2B^2)^2}}{1+m^2B^2}\,.
\end{align}
The \emph{extremal} case ${\M=0}$ is obtained for the rotation ${a_{\rm e}=m/(1+m^2 B^2)}$, and such horizon is at
    \begin{align}\label{extremal}
        r_{\rm e}=\dfrac{m}{1+m^2B^2}= a_{\rm e}\,.
    \end{align}
    Notice that it is smaller then  \label{fig1} for the extremally rotating Kerr black hole. Also, ${\M=0}$ implies ${P=1}$.

\item In the case ${m=0}$, the horizons are at
\begin{align}\label{horiz-m=0case}
    r_{\pm} = \pm a\, \sqrt{(a^2+e^2)/(e^4 B^2 -a^2)}\,.
\end{align}
It necessarily requires ${e\ne0}$, ${B\ne0}$, and ${a<e^2B}$. Otherwise the spacetime represents a specific naked singularity without horizons. Such particular exact type~D solutions are generalizations of massless Reissner-Nordstr\"om and massless Kerr-Newman spacetimes which also have no horizons, cf. \eqref{Psi2-B=0}.

\end{itemize}

\subsection{Regularity of the axes}

The Kerr-Newman-Bertotti-Robinson black hole \eqref{metr} has a spherical topology of horizons, with the poles at ${\theta=0}$ and ${\theta=\pi}$. In fact, they identify the axes of symmetry in the whole spacetime, and these can be made regular by choosing a specific value of the conicity parameter $C$, fixing the range of the angular coordinate to
\begin{align}
    \varphi\in [0,2\pi C)\,.
\end{align}
This special value of the constant~$C$ is defined by the usual condition, namely that the ratio of the proper length of a small circumference ${\int_{0}^{2\pi C}\sqrt{g_{\varphi\varphi}}\,\dd \varphi}$ and its radius ${\int\sqrt{g_{\theta\theta}}\,\dd \theta}$ is $2\pi$. This yields the condition ${C\,P(\theta=0, \hbox{ or } \theta=\pi)=1}$. Because all the metric functions in the metric (\ref{metr}) are symmetric with respect to the equatorial plane ${\theta=\pi/2}$, on \emph{both} poles these ratios are the same, and can be set to $2\pi$ by the \emph{unique} choice
\begin{align}\label{conicity}
    C=\frac{1}{1+B^2\M^2}\,.
\end{align}
In the ${B=0}$ case we obtain ${C=1}$, meaning that for the Kerr-Newman spacetime  both the axes are regular with the usual range ${\varphi\in[0,2\pi)}$. In the Kerr-BR$_s$ case (${e=e_s}$, ${k=0}$) we obtain $C=1/\big[\,1+B^2\,[m^2/(1-a^2 B^2)-a^2 ]\,\big]$. In view of \eqref{mp}, this is exactly Eq.~(77) presented in \cite{PodolskyOvcharenko:2025}. In the genuine Kerr-BR$_0$ case that has no charge (${e=0}$, ${k=-mB}$), the conicity becomes
\begin{align}
    C=\frac{1}{(1+m^2B^2)\big[1-a^2B^2(1+m^2B^2)\big]}\,.
\end{align}
For ${a=0}$ we get the Schwarzschild-BR black hole, regular for ${C=1/(1+m^2B^2)}$, see Eq.~(101) in~\cite{OvcharenkoPodolsky:2026a}.

\subsection{Electric and magnetic charges}
\label{charges}

It is also crucial to analyze the electromagnetic field, in particular the physical charges of the black hole. To this end, we first calculate the electric and magnetic fluxes through the outer horizon of the black hole, given by the  Faraday 2-form $\mathbf{F}=\dd \mathbf{A}$, where $\mathbf{A}$ is given by \eqref{A_vec-explicitly}, and its Hodge dual $\tilde{\mathbf{F}}$. Using these 2-forms we integrate their flux through the horizon, and thus by using the Gauss law we determine the physical electric and magnetic charges,
\begin{align}
    q_e \equiv & \textstyle{-\frac{1}{4\pi}\int_0^{2\pi C}\!\int_{-1}^{1}\tilde{F}_{x\varphi}\,\dd x\,\dd\, \varphi}\,,\nonumber\\
    q_m \equiv & \textstyle{-\frac{1}{4\pi}\int_0^{2\pi C}\!\int_{-1}^{1}F_{x\varphi}\,\dd x\,\dd \varphi}\,.
\end{align}
Direct integrations give us (for more details see \cite{supp_mat})
\begin{align}\label{qe,qm}
    q_e=C\, e\,,  \qquad q_m=0\,.
\end{align}
The magnetic charge is zero, and the \emph{black hole has only electric charge}. Moreover, this electric charge is given simply by the parameter $e$ multiplied by the conicity~$C$. Miraculously, this nice relation holds \emph{for any value of the rotation parameter $a$, and for any value of the external magnetic field $B$}. This justifies using $e$ as the proper charge parameter in the metric functions \eqref{Delta_eq}.

\subsection{Thermodynamical properties}
\label{sec:thermodynamics}

Finally, we calculate basic thermodynamic quantities, namely the \emph{entropy}~${S =\tfrac{1}{4}\,{\cal A}}$ and \emph{temperature}~${T = \tfrac{1}{2\pi}\,\kappa}$ of the black hole horizons located at ${r=r_h}$, where ${\cal A}$~is the horizon area, and $\kappa$~is the surface gravity \cite{Wald:book1984}.

The horizon area is obtained by integrating the angular coordinates of the metric \eqref{metr} for fixed $t$ and ${r=r_h}$, namely ${\mathcal{A}(r_h) = \int_0^{2\pi C}\!\int_0^\pi \sqrt{g_{\theta \theta}\, g_{\varphi \varphi}}\,\,\dd \theta \, \dd \varphi}$. Using the fact that ${{Q}(r_h)=0}$ on the horizon, we get ${\mathcal{A} = 2\pi C\,\big(r_h^2+a^2\big) \int_0^\pi \Omega^{-2}(r_h)\,\sin\theta\,\dd \theta}$. Moreover, $\Omega^2$ evaluated on the horizon is a \emph{constant} ${\Omega^2(r_h) = I(r_h) = (1 + \con\,B\,r_h)^2+B^2 r_h^2}$, so that
\begin{align}\label{Ah}
    \mathcal{A}_h=4\pi C\, \dfrac{r_h^2+a^2}{(1 + \con\,B\,r_h)^2+B^2 r_h^2}\,.
\end{align}
Recall that $r_h$ is either $r_+$ or $r_-$ given by \eqref{general-horizons}. It generalizes the usual expression ${\mathcal{A}_h=4\pi \,(r_h^2+a^2)}$ for the Kerr-Newman black hole, which is recovered by setting ${B=0}$. For ${k=0}$ we obtain the previous result for the Kerr-BR$_s$ black hole (see Eq.~(33) in \cite{PodolskyOvcharenko:2025}). The genuine Kerr-BR$_0$ spacetime without charge (${e=0}$) has ${\con = - m B}$.

The \emph{surface gravity}~$\kappa$ of the horizon at~$r_h$ is defined as the acceleration of the null normal vector~${\xi^a=\partial_t+\omega_h\, \partial_{\varphi}}$,  with ${\omega_h=-g_{t\varphi}/g_{\varphi\varphi}}$ (the horizon generator), using the relation ${\xi_{a;b}\,\xi^b  = \kappa\, \xi_a}$. It can be calculated by the formula ${\kappa_h = \tfrac{1}{2} Q'(r_h)/(r_h^2+a^2)}$, see \cite{PodolskyVratny:2021}. In view of
\eqref{Delta_eq}, on the horizon we have ${Q'(r_h)=I(r_h)\,\Delta'(r_h)}$, and thus
\begin{align}\label{kappa-h}
    \kappa_h = \dfrac{(1 + \con\,B\,r_h)^2+B^2 r_h^2}{r_h^2+a^2}\,
       \Big[ \, m - (1+\con^2)\,\frac{a^2}{r_h} \,\Big].
\end{align}
For ${B=0}$, ${\kappa_h = \big[m-(a^2+e^2)/r_h\big]/(r_h^2+a^2)}$, equivalent to
the formula  ${\kappa_+ = \tfrac{1}{2}(r_+-r_-)/(r_+^2+a^2)}$ for the Kerr-Newman black hole. Recall also that when the condition \eqref{extremality} holds, there is an extremal horizon located at ${r_{\rm e}=a^2(1+k^2)/m}$. In this case, \eqref{kappa-h} gives ${\kappa_{\rm e}=0}$, and such \emph{extreme black holes} thus have \emph{zero temperature}~$T$.

By combining \eqref{kappa-h} and \eqref{Ah} we obtain
\begin{equation}
2\,T S \equiv \frac{1}{4\pi}\,\kappa_{h}\,\mathcal{A}_h
  =  C \Big[ \, m - (1+\con^2)\,\frac{a^2}{r_h} \,\Big].
\label{Smarr}
\end{equation}
Using \eqref{general-horizons}, at $r_+$ this can be equivalently rewritten as
\begin{equation}
2\,(T S)_+  =  C \M \,,
\label{Smarr-KN-BR}
\end{equation}
in which the conicity~$C$ is given by \eqref{conicity}, and the parameter $\M$ is given by expression
\eqref{M_eq}. For ${B=0}$ we recover ${2(T S)_+ = \sqrt{m^2-a^2-e^2}}$, while for the genuine uncharged Kerr-BR$_0$ black hole (${e=0}$) we get
\begin{equation}
2\,(T S)_+  = \frac{\sqrt{m^2-a^2(1+m^2B^2)^2}}{1+B^2\,[m^2-a^2(1+B^2 m^2)^2]}\,.
\label{Smarr-Kerr-BR0}
\end{equation}
For the Schwarzschild-BR black hole (${a=0}$) it reduces to ${2\,(T S)_+ = m/(1+B^2m^2)}$, cf.~\cite{PodolskyOvcharenko:2025}. Thermodynamics of the genuine Kerr-BR black holes is investigated in~\cite{KubiznakOvcharenkoPodolsky:2026b}.

\section{Conclusions}

We have introduced and investigated an explicit type~D exact solution to the Einstein-Maxwell equations that describes arbitrarily rotating and charged black hole immersed in an external magnetic field. It can be called the Kerr-Newman-Bertotti-Robinson (KN-BR) class because for zero magnetic field (${B=0}$) it gives the usual Kerr-Newman metric, while for zero mass (${m=0}$) the Bertotti-Robinson universe with a uniform magnetic field is recovered. Moreover, its linearization in $B$ reduces to the Wald spacetime \cite{Wald:1974}.

Our solution, obtained as the special subcase of spacetimes \cite{OvcharenkoPodolsky:2025} with a non-aligned electromagnetic field, has a surprisingly simple form \eqref{metr}--\eqref{A_vec-explicitly} with the metric functions factorized to just quadratic expressions in the coordinates $r$ and $\cos\theta$. This is in striking contrast to the famous Kerr-Newman-Melvin  (Ernst-Wild) class of black holes \cite{Ernst1976_2, Ernst1976_3}, cf. its explicit two-pages-long metric written  as Eqs.~(B.4)--(B.18) in \cite{GibbonsMujtabaPope:2013}.

Also, the total electric charge $q_e$ of the Kerr-Newman-Melvin black hole is a qubic polynomial of the seed Kerr-Newman charge~$e$, namely ${q_e= e +2amB -\frac{1}{4}B^2e^3}$, see~Eq.~(3.3) in \cite{GibbonsMujtabaPope:2013}. In the new Kerr-Newman-Bertotti-Robinson case this is simply given by~${q_e=C\,e}$, where $C$ is the conicity of the axis. This is of a great help if one wants to identify the genuine \emph{neutral} rotating Kerr black hole in the external magnetic field. In the Kerr-Newman-Melvin  case it is necessary to solve the qubic equation ${q_e=0}$ to determine the corresponding seed parameter $e$, while in the KN-BR case we simply set ${e=0}$.

This nice property enabled us to prove that the Kerr-BR$_s$ solution presented in \cite{PodolskyOvcharenko:2025} describes a black hole that \emph{is electrically charged}. Actually, its total charge has a special value ${e_s = m\,a\,B/\sqrt{1-a^2B^2}}$, see~\eqref{especial}. The \emph{genuine} Kerr-like black hole \emph{without any charge} in the Bertotti-Robinson universe is simply given by ${e=0}$, and we denoted it as Kerr-BR$_0$. These two subfamilies of a generic KN-BR class are distinct, but they \emph{both} reduce to the neutral Kerr black hole for ${B=0}$.

The new KN-BR metric has also other  aspects favourable for investigation of physical properties, such as the position of horizons, curvature and topological singularities, thermodynamics, or geodesics. We thus hope that our metric will be useful for further studies in mathematical relativity and astrophysics.

\section*{Acknowledgments}

This work has been supported by the Czech Science Foundation Grant No.~GA\v{C}R 26-22381S, and by the Charles University Grant No.~GAUK 260325.

\end{document}